\documentclass[aps,prx,twocolumn,amsmath,amssymb,superscriptaddress,floatfix]{revtex4-2}

\usepackage{mathtools}
\usepackage{multirow}
\usepackage{xcolor}
\usepackage{bm}
\usepackage{amsfonts,amssymb,amsmath}
\usepackage{graphicx,dcolumn,bm,xcolor,braket,slashed}
\usepackage{times} 
\usepackage{comment}
\usepackage{array}
\usepackage{textcomp}
\usepackage[normalem]{ulem}
\usepackage{dsfont}

\newcommand{\be}{\begin{eqnarray}}
\newcommand{\ee}{\end{eqnarray}}
\newcommand{\bbm}{\begin{bmatrix}}
\newcommand{\ebm}{\end{bmatrix}}
\newcommand{\bpm}{\begin{pmatrix}}
\newcommand{\epm}{\end{pmatrix}}

\newcommand{\blue}[1]{{\color{blue} #1}}

\usepackage[toc,page,header]{appendix}
\usepackage{minitoc}

\begin{document}

\doparttoc 
\faketableofcontents 

\part{} 

\title{Thermoelectric Transport Driven by Quantum Distance}

\author{Chang-geun Oh}
\affiliation{Department of Applied Physics, The University of Tokyo, Tokyo 113-8656, Japan}
\author{Kun Woo Kim}
\email{kunx@cau.ac.kr}
\affiliation{Department of Physics, Chung-Ang University, 06974 Seoul, Republic of Korea}
\author{Jun-Won Rhim}
\email{jwrhim@ajou.ac.kr}
\affiliation{Department of Physics, Ajou University, Suwon 16499, Republic of Korea}
\affiliation{Research Center for Novel Epitaxial Quantum Architectures, Department of Physics, Seoul National University, Seoul 08826, Republic of Korea}

\begin{abstract}
%
%
The geometric characteristics of Bloch wave functions play a crucial role in electronic transport properties.
We show that the thermoelectric performance of materials is governed by the geometric structure of Bloch wave functions within the framework of the Boltzmann equation.
The essential geometric notion is the Hilbert-Schmidt quantum distance, measuring the resemblance between two quantum states.
We establish a geometric characterization of the scattering rate by extending the concept of quantum distance between two states in momentum space at a distance.
Employing isotropic quadratic band touching semimetals, where one can concentrate on the role of quantum geometric effects other than the Berry curvature, we find that the response functions for electrical quantum transport and, therefore, the thermoelectric power factor can be succinctly expressed in terms of the maximum quantum distance, $d_\mathrm{max}$.
Specifically, when $d_\mathrm{max}$ reaches one, the power factor doubles compared to the case with trivial geometry ($d_\mathrm{max}=0$). 
Our finding highlights the significance of quantum geometry in improving the performance of thermoelectric devices.
\end{abstract}

\maketitle

\textit{Introduction.---}
In modern solid-state physics, the Berry curvature~\cite{berry1984quantal}, one of the geometric properties of a Bloch wavefunction~\cite{shapere1989geometric}, has played an essential role in understanding various anomalous transport phenomena~\cite{xiao2010berry,nagaosa2010anomalous,jungwirth2002anomalous,nayak2016large} and the topological nature of solids~\cite{thouless1982quantized,kohmoto1985topological,haldane1988model}.
The Berry curvature acts as an emergent magnetic field in the semiclassical equation of motion of solids~\cite{xiao2010berry}, causing wave packets to move with an anomalous velocity proportional to the Berry curvature. 
This effect leads to various Hall-like effects, such as the anomalous Hall, valley Hall, spin Hall, and anomalous phonon Hall effects~\cite{hirsch1999spin,sinova2015spin,jungwirth2002anomalous,shimazaki2015generation,qin2012berry,xiao2007valley,yao2008valley,xiao2012coupled,li2013unconventional,kormanyos2018tunable,mak2014valley}.
Additionally, the first Chern number, which can be calculated by integrating the Berry curvature over the Brillouin zone, provides insights into electronic structures of the edge states in the the time-reversal broken systems, directly relating to the quantization of Hall conductivity quantization~\cite{thouless1982quantized}.

However, the Berry curvature is not the only geometric attribute of the Bloch wavefunction. 
The geometric description also involves the Hilbert-Schmidt quantum distance, which leads to the quantum geometric tensor. This tensor's real and imaginary parts correspond to the quantum metric and Berry curvature, respectively~\cite{provost1980riemannian,shapere1989geometric,cheng2010quantum,hwang2021geometric}.
Unlike the Berry curvature, the physical implications of the quantum geometric tensor and quantum distance have only recently been explored in contexts such as charge and spin Hall effects under an inhomogeneous electric field~\cite{lapa2019semiclassical,wang2023quantum,kozii2021intrinsic,zhang2022geometric}, current noise~\cite{neupert2013measuring}, electron-phonon coupling~\cite{yu2023nontrivial}, superfluid weight~\cite{peotta2015superfluidity,liang2017band,herzog2022superfluid,peotta2023quantum}, various magnetic responses~\cite{piechon2016geometric,rhim2020quantum,jung2024quantum,oh2024revisiting}, and bulk-boundary correspondence~\cite{oh2022bulk,kim2023general}.

\begin{figure}[t]
\includegraphics[width=85mm]{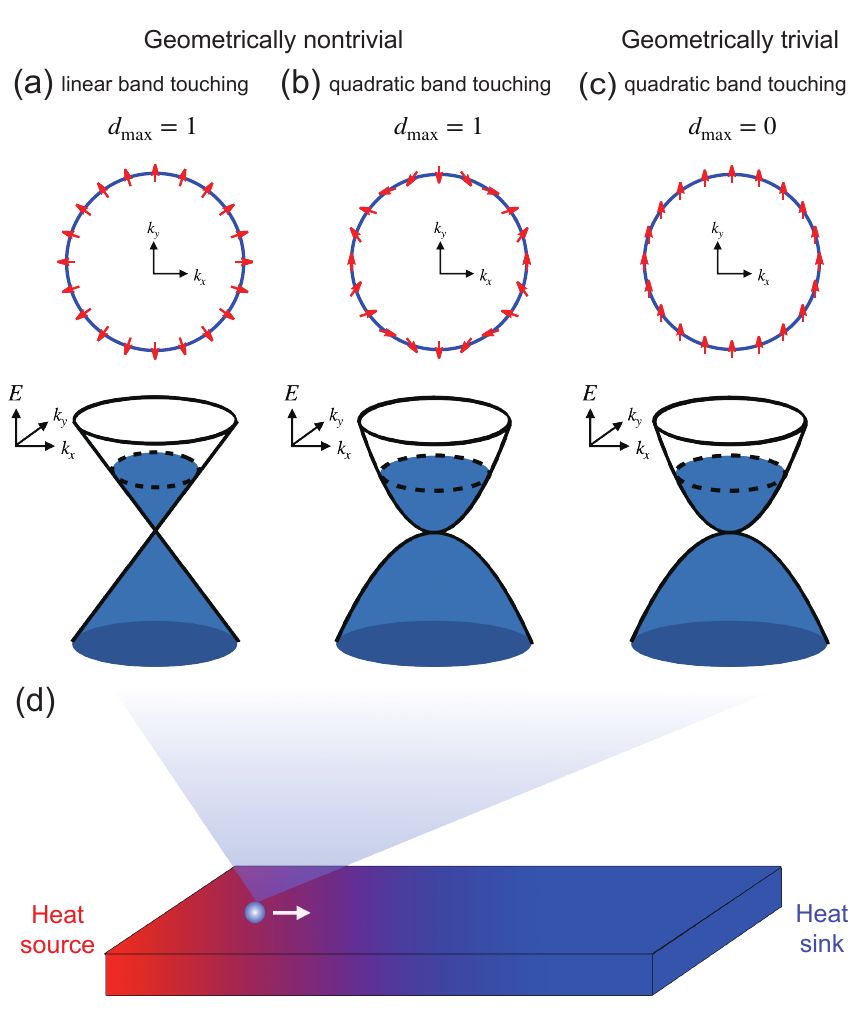} 
\caption{\label{fig1}
Schematic representations of the low-energy band structure and of the pseudospins $(s_y(\bm{k}),s_z(\bm{k}))$ for (a) linear band touching system $d_\mathrm{max}=1$, (b) quadratic band touching with $d_\mathrm{max}=1$, and (c) quadratic band touching with $d_\mathrm{max}=0$, respectively. (d) Schematic illustration of mesoscopic thermoelectric device under a temperature gradient. }
\end{figure}

In this paper, we explore the relationship between quantum distance and thermoelectric properties. 
{We begin by deriving the scattering probability and disorder-averaged transport scattering rate for a general two-band Hamiltonian, showing that these quantities can be expressed in terms of the quantum distance and pseudospins of the Bloch eigenvectors. 
Then, we show that those formulas reduce to a simple form depending on the maximum quantum distance in the quadratic band-touching cases, where the Berry curvature vanishes.
Namely, in quadratic band-touching semimetals, we demonstrate that the thermoelectric power factor is characterized by the maximum quantum distance between Bloch eigenvectors near the touching point. 
The schematics are illustrated in Fig. \ref{fig1}.
Finally, we clarify the impact of quantum geometry on thermal transport behavior by considering a semimetal model, in which the maximum quantum distance can be varied while the band structure remains the same.}

\textit{The figure of merit in {thermoelectrics}.---}
A temperature gradient in a system induces the diffusion of charged particles, resulting in an electrical current:
\begin{align}
    j^c = -\sigma \nabla V - \sigma S \nabla T,
\end{align}
where the Seebeck coefficient $S$ converts the temperature gradient {$\nabla T$} to an effective electric field. 
For a system in thermal equilibrium with no electric current ($j^c=0$) and no electric potential ($\nabla V=0$), introducing a small temperature gradient  $\delta(\nabla T)$ induces an electric current given by $\delta j^c = -\sigma S \delta (\nabla T)$. 
Consequently, the change in the electric field across the device is $\delta (\nabla V) = -S \delta (\nabla T)$, as defined by the Seebeck coefficient $S=-dV/dT$.
Thus, the electric power generated per unit length by the temperature gradient is:
\begin{align}
    \delta P_e = \delta j^c \delta (\nabla V) = \sigma S \delta (\nabla T) S \delta (\nabla T). 
\end{align}
{Simultaneously, the rate of entropy production is given by}
\begin{align}
    \frac{d}{dt}\Sigma = \frac{1}{T} \kappa \delta (\nabla T). 
\end{align}
where $\kappa$ is thermal conductivity, $\kappa \delta (\nabla T) $ is the heat flux $\dot Q$, and entropy change $\dot \Sigma =\dot Q/T$.
The figure of merit $ZT$ is defined as the ratio of the electric power to the entropy production per $\delta (\nabla T)$: 
\begin{align}
    ZT = \frac{1}{\delta (\nabla T)}\frac{\delta P_e}{ \dot \Sigma} = \frac{\sigma S^2 T}{\kappa}. 
\end{align}
In semi-metallic systems at room temperature, the thermal conductivity is dominated by phonon, and the power factor (PF), $PF=\sigma S^2$, essentially determines the $ZT$. We show how it is closely connected to the distribution of quantum distances on the Fermi surface. 

Under the relaxation time approximation, the tilting of the Fermi surface due to a temperature gradient and electric field {($\bm E$)} is described by: 
\begin{align}
f(n,\bm{k},\bm r)-f_0 =   \frac{(\hbar/\tau_{n\bm{k}})^{-1}}{2+2\cosh\xi} \left[e\bm E + k_B T \nabla_{\bm r} \xi \right] \cdot \nabla_{\bm k}\xi,
\label{eq:distribution}
\end{align}
where $\xi\equiv (\epsilon_{n\bm{k}}-\mu)/k_BT$, \,
$\tau_{n\bm{k}}$ is the mean scattering time of $n$-th band, and $f_0=(\mathrm{exp}[(\epsilon_{n\bm{k}}-\mu)/k_B T]+1)^{-1}$ is the Fermi-Dirac distribution. The charge and thermal current densities are 
\begin{align}
    \bm j^c &= \frac 1 V \sum_{n,\bm k} \bm e v_{n\bm k} \,f(n,\bm k,\bm r), \nonumber \\
    \bm j^{th} &= \frac 1 V \sum_{n,\bm k} (\epsilon_{n\bm k}-\mu) \bm v_{n\bm k} \,f(n,\bm k,\bm r),\label{eq:currents}
\end{align}
where $V$ is {the volume of the system}. 
The power factor and the figure of merit associated with electronic transport can be directly computed using \eqref{eq:distribution} and \eqref{eq:currents}.

\textit{Quantum distance, pseudospin, and transport scattering rate.---}
Consider an $N$-dimensional Hilbert space described by the eigenstates $\ket{\psi_n (\bm{\Lambda})}$, that smoothly depend on a set of real parameters $\bm{\Lambda} =(\Lambda_1,\Lambda_2,...)$, where $n \in \{1,...,N \}$ denotes the band index. 
The Hilbert-Schmidt quantum distance for $n$th band is defined as follows~\cite{provost1980riemannian}:
\begin{eqnarray}
        d_{\text{HS},n}^2 (\bold{\Lambda} , \bold{\Lambda'}) = 1 - |\braket{\psi_n({\bold{\Lambda})}|\psi_n({\bold{\Lambda'})}}|^2,
\end{eqnarray}
which quantifies the distance between quantum states. 
This distance reaches unity when the states are orthogonal and zero when parallel up to a global phase. 
In systems with discrete translation symmetry, the crystal momentum $\bm k$ takes the role of $\bm \Lambda$. 

In a two-band system
the generic form of the Hamiltonian is given by 
\begin{eqnarray}
    H_{\bm{k}}= h_{0,\bm{k}}{\sigma_0} + \bm{h}_{\bm{k}} \cdot \bm{\sigma},
\end{eqnarray}
where $\bm{\sigma}=(\sigma_x,\sigma_y,\sigma_z)$ are the Pauli matrices in the (pseudo-)spin basis and $h_i$ is a real-valued function of momentum.
{The pseudospin textures visualize the geometric properties of the Bloch eigenvectors, as illustrated in Fig.~\ref{fig1}(a) to (c).
In graphene, the texture can be described by designating two of its sublattices as the pseudospin components. 
Near the linear band-touching of graphene, the average pseudospin vectors exhibit a chiral structure with winding number one, as illustrated in Fig.~\ref{fig1}(a).
This winding structure is closely linked to unconventional transport properties, such as the half-integer quantum Hall effect~\cite{novoselov2005two,zhang2005experimental}, the absence of back-scattering processes \cite{ando1998impurity,ando2005theory}, Klein tunneling \cite{katsnelson2006chiral,young2009quantum}, and the weak anti-localization phenomena \cite{mccann2006weak,tikhonenko2009transition}.

Recently, it was reported~\cite{rhim2020quantum} that semimetals with a quadratic band touching point can exhibit a canted structure in their pseudospin texture without a quantized winding number.
Focusing on the upper band, the pseudospin on the Fermi surface is $\bm{s}_{\bm k}=\bra{+\bm k} \hat S \ket{+\bm k}=\bm{h}/|\bm{h}|$ 
\footnote{The direction of pseudospin is defined as $\bm s(\bm k) = \bra{u_+(\bm k)} \bm \sigma \ket{u_+(\bm k)}=-\bra{u_-(\bm k)} \bm \sigma \ket{u_-(\bm k)}$. }. 
The quantum distance between states can be read out from the pseudospin structure illustrated in Fig.~\ref{fig1}(a) to (c) through the following relation
~\footnote{The fidelity $|\langle k|k' \rangle|^2=\langle k' | k \rangle \langle k | k' \rangle=\bra {k'} \frac 1 2 + \frac 1 2 \vec s_{k} \cdot \vec \sigma \ket {k'}=\frac 1 2 + \frac 1 2 \vec s_{k'} \cdot \vec s_k$, where $\vec s_k = \bra k \vec \sigma \ket k$. The relation can be verified using $\ket k=\bpm \cos \frac {\theta_k} 2 e^{-i\phi_k} & \sin \frac {\theta_k} 2 \epm^T$ and $\vec s_k=(\sin\theta_k \cos\phi_k, \sin\theta_k \sin\phi_k, \cos\theta_k)$.}
:
    \begin{eqnarray}
        &&d_{\bm{kk'}}^2=\frac{1}{2}\left[1-\bm{s}_{\bm{k}}\cdot \bm{s}_{\bm{k'}}\right],
    \end{eqnarray}
indicating that $d_{\bm{kk'}}=0$ ($d_{\bm{kk'}}=1$) when pseudospin vectors are (anti-)parallel.

The transport scattering rate for a state $\ket{\bm{k}}\equiv \ket{+\bm k}$ in the upper band, due to a perturbative potential $U_{\text{pert}}(\bm r)$ that breaks the translation symmetry, typically from material imperfections, is given by: 
    \begin{eqnarray}
        \frac{1}{\tau_{{\bm{k}}}}=\sum_{\bm{k'}\in\text{FS}} W_{\bm{k'}{\bm{k}}}(1-\mathrm{cos}\theta_{\bm{k'}{\bm{k}}}), \label{eq:scat_rate}
    \end{eqnarray}
where $W_{\bm{k'k}}$ represents the scattering probability per unit time, defined as:
\begin{align}
W_{\bm{k'k}}=\frac{2\pi}{\hbar}|\braket{{\bm{k}}|U_\text{pert}(\bm{r})|\bm{k'}}|^2 \delta(\epsilon_{\bm{k'}}-\epsilon_{{\bm{k}}}),
\end{align}
and {$\theta_{{kk'}}\text{=}\cos^{-1} (\bm \hat {\bm k}\cdot \hat {\bm k'})$} is the scattering angle, with $\hat{\bm{k}}\,\text{=}\,\bm {k}/|\bm{k}|$. 
The generic form of the scattering potential is expressed as $U_\text{pert}=\sum_{i=0,x,y,z} U_i(\bm{r})\sigma_i$, where it can cause both momentum transfer and a (pseudo-)spin rotation.
Employing the Fermi Golden rule, the scattering probability is 
\begin{align}
    W_{\bm{k'k}} &=  \frac{2\pi}{\hbar}\Bigl[ (\bm v_{\bm q}\cdot \bm v_{\text{-}\bm q}- |v_{\bm q,0}|^2)\frac{1-\bm s_{\bm k} \cdot \bm s_{\bm{k'}}}{2} \bigr.\nonumber\\ 
    &+ \bigl. \sum_{i,j=0,x,y,z} (v_{\bm q,i}\bm s_{\bm k,i}) (v_{\text{-}\bm q,j}\bm s_{\bm k',j}) \Bigr] \delta(\epsilon_{\bm{k'}}-\epsilon_{\bm k}), \label{eq:scat_prob}
\end{align}
{where $\bm v_{\bm q}$ is a four-component vector with components given by $v_{\bm q,i}=V^{-1}\int d^2\bm{r}\,U_i(\bm{r})e^{-i\bm q\cdot\bm{r}}$ ($i=0,x,y,z$)}
, and $\bm q=\bm k'-\bm k$ is the momentum transfer. Here, we denote $\bm s_{\bm k,i}\equiv (\bm s_{\bm k})_i$ and $(\bm s_{\bm k})_{i=0}=1$. 
%
This formulation shows that the scattering probability depends on the pseudospins of the eigenstates and the properties of the extrinsic scattering potential. The relation highlights the role of both intrinsic material properties and external perturbations in determining the transport characteristics. 
The Fermi Golden rule including the higher order corrections is in supplemental materials (SM), and one can verify that scattering probability is a function of the pseudospin and scattering potential, even when higher order corrections are included.

The distribution of the scattering potential is information specific to material and growth processes.
To derive a generic relation between the transport scattering rate and the quantum distance, we assume a Gaussian distribution for the scattering potentials. 
Upon averaging over disorder, the crossing terms containing pairs of uncorrelated scattering potential ($\langle v_{\bm q,i}v_{{-}\bm q,j}\rangle_{i\neq j}$) vanish.
Consequently, the transport scattering rate is simplified to
\begin{align}
     \biggl\langle \frac{1}{\tau_{\bm k}} \biggr \rangle 
     = \frac{2\pi}{\hbar}\sum_{\bm{k'}\in \text{FS}} &\Bigl[ \gamma_0^2 (1-d_{\bm{kk'}}^2) +\frac{|\bm{\gamma}|^2}{3} (1+d^2_{\bm{kk'}})\Bigr] \nonumber\\
     &\times (1-\cos\theta_{kk'}), 
     \label{eq:discatrate}
\end{align}
where $\gamma_0^2=\langle |v_{q,0}|^2\rangle$ and $\gamma_i^2=\langle |v_{q,i}|^2\rangle$ are the averaged strengths of the scalar and vector components of the scattering potential, respectively.
This result explicitly shows that the distribution of the quantum distance $d_{\bm k\bm k'}$ across the Fermi surface is the critical factor in determining the transport coefficients. 
Specifically, the presence of nontrivial quantum geometry in a system can reduce (increase) the scattering rate when the scattering originates from spin-independent (spin-dependent) potentials.
In the following we demonstrate this relationship using a 2-dimensional isotropic model.

\textit{Application to 2D isotropic band model.---}
We consider a 2D isotropic system, where not only the magnitude of the Fermi velocity $v_F=|\partial_{\bm k}\epsilon_{\bm k}|$ remains constant on the Fermi surface, but also the pseudospin vector rotates uniformly with a constant angular velocity with respect to $\theta_k$ on the Fermi surface. 
That is, we consider a system with spatial rotation combined with spin rotation symmetry in the Hamiltonian, $[\hat H_0, \hat U_{\phi, W}]=0$, where 
\begin{align}
    \hat U_{\phi,W} = e^{i\phi \hat L_z/\hbar}\otimes e^{i W\phi (\hat s_0 \cdot \bm \sigma)/
    \hbar},\label{eq:U_sym}
\end{align}
implying that when the momentum angle $\theta_k$ changes by $\phi$, the pseudospin vector rotates around $\hat s_0$ by an angle $W\phi$. 
The cases illustrated in Fig.~\ref{fig1}(a), (b) and (c) correspond to $W=1,~2$, and $0$, respectively. 
The trajectory of the pseudospin vector can be parameterized by $\theta_k$: 
\begin{align}
    \bm s_{\bm k }=\bm s_{0} + \bm s_{\perp\alpha} \cos(W\theta_{\bm k}) + \bm s_{\perp\beta}\sin(W\theta_{\bm k}),
\end{align}
where {$\bm s_{0}$, $\bm s_{\perp\alpha}$, and $\bm s_{\perp\beta}$} are mutually orthogonal, $|\bm s_0|\text{=}\sqrt{1-d_\mathrm{max}^2}$, and $|\bm s_{\perp \alpha}|\text{=}|\bm s_{\perp \beta}|\text{=}d_\mathrm{max}$. 
Then, the quantum distance is given by
\begin{align}
    d^2_{\bm{kk'}} = d_\mathrm{max}^2\frac{1-\cos W\theta_{kk'}}{2}
\end{align}
where $d_\mathrm{max}=\text{max}_{\bm k,\bm k'\in \text{FS}}\left[d_{\bm{kk'}}\right]$ represents the maximum quantum distance between all the possible pairs of Bloch eigenvectors on the Fermi surface.
The scattering probability in Eq.~(\ref{eq:scat_prob}) is strongly influenced by $W$ and $d_\mathrm{max}$, as shown in Figs.~\ref{fig2}(a) and (b). 
Averaging over the disorder ensemble, the transport scattering rate from Eq.~(\ref{eq:discatrate}) becomes 
\begin{align}
    \biggl\langle \frac{1}{\tau^{\text{iso}}_{\bm k}} \biggr \rangle =&\frac{2\pi}{\hbar}\frac{\rho(\epsilon_F)}{2} \bigg[\gamma_0^2 (2-d_\mathrm{max}^2) + 
    \frac{|\bm \gamma|^2}{3} (2+d_\mathrm{max}^2)\bigg]  \nonumber \\
    &+d_\mathrm{max}^2\frac{2\pi}{\hbar}\rho(\epsilon_F)\frac{|\bm \gamma|^2-3\gamma_0^2}{12}\delta_{W,1}, \label{eq:tau_iso}
\end{align}
where $\rho(\epsilon_F)$ is the density of states at the Fermi level and the last term ($\sim \delta_{W,1}$) is nonzero for $W{=}1$. 
This expression shows that the transport scattering rate depends solely on the maximum quantum distance between possible scattering states and the disorder strength ($\gamma_{0}$ and $|\bm \gamma|$). 
When $\gamma_0^2 > |\bm{\gamma}|^2/3$ ($\gamma_0^2 < |\bm{\gamma}|^2/3$), increasing $d_\mathrm{max}$ decreases (increases) the scattering rate, {as shown in Figs.~\ref{fig2}(a) and (b).} 
Notably, the winding number $W$ only impacts the scattering rate for $W=1$.
%
Since the power factor is proportional to the inverse of the transport scattering rate, $\text{PF} \propto \langle 1/\tau^{\text{iso}}\rangle$, the maximum quantum distance plays a crucial role in determining the thermoelectric performance of the isotropic model. 

\begin{figure}[t]
\includegraphics[width=85mm]{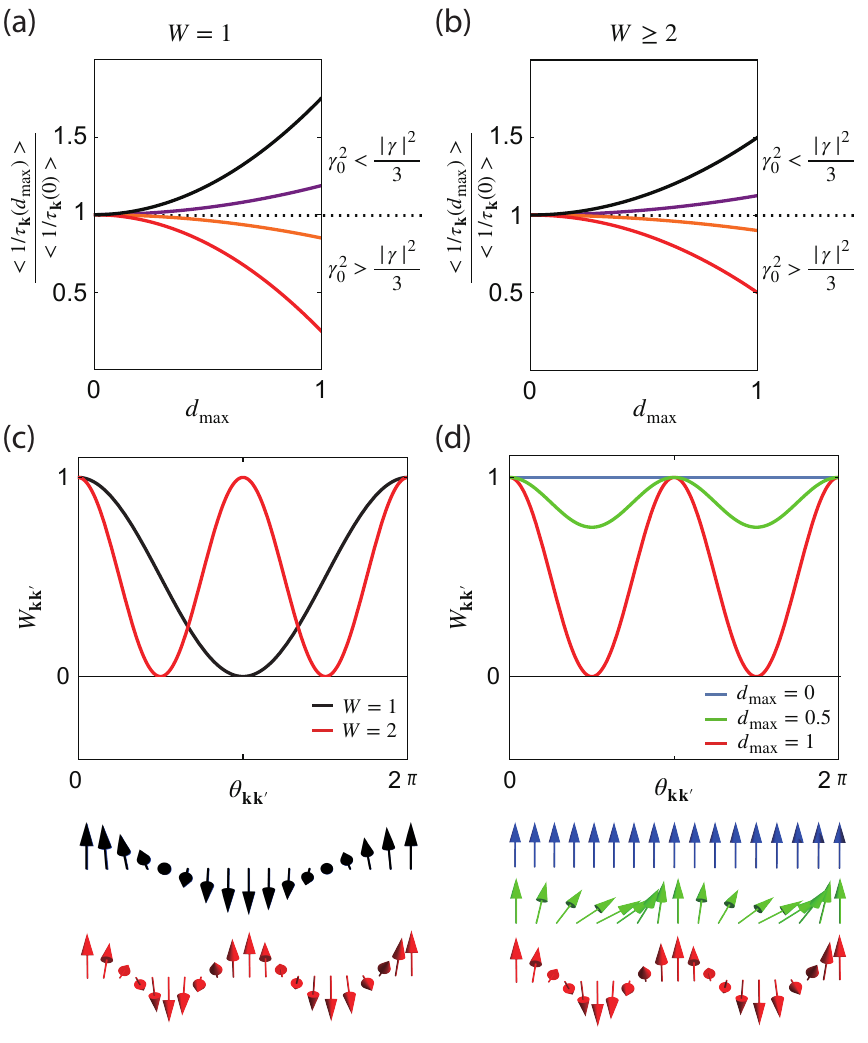} 
\caption{\label{fig2}
(a-b) The maximum quantum distance $d_\mathrm{max}$ dependence of scattering rate $1/\tau_{\bm{k}}$ with (a) $W=1$ and (b) $W\geq2$. The red, orange, purple, and black lines represents $(\gamma_0^2,|\bm{\gamma}|^2)=c(1,0), c(1,2), c(1,5)$ and $c(0,1)$, where $c$ is a positive constant. (c) Scattering probability $W_{\bm{kk'}}$ with $|\bm{k'}|=|{\bm{k}}|$ and pseudospin structure as a function of scattering angle $\theta_{\bm{kk'}}$ for $W=1$ (black) and $2$ (red). 
(d) Scattering probability $W_{\bm{kk'}}$ with $|\bm{k'}|=|{\bm{k}}|$ and pseudospin structure as a function of scattering angle $\theta_{\bm{kk'}}$ for the isotropic quadratic band touching. The blue, green, and red plots represent $d_\mathrm{max}=0$, 0.5 and 1, respectively. In (c-d), we consider $\bm{\gamma}=\bm{0}$ and $2\pi\gamma_0^2/\hbar =1$.} 
\end{figure}


\textit{Application to 2D quadratic isotropic band touching model.---}
The quadratic isotropic band-touching semimetals with symmetry $\hat U_{\phi,W=2}$ provide an ideal platform to study the quantum distance and quantum metric. This model allows for the manipulation of the geometric structure of wavefunctions while keeping the band structure constant. Additionally, in this model, the Berry curvature is absent~\cite{rhim2020quantum,rhim2021singular}.
The model is characterized by three parameters: the mass of the upper (lower) band $m_+ (m_-)$, and the maximum quantum distance $d_\mathrm{max}$. 
The Hamiltonian is given by $\mathcal{H}_{0}({\bm{k}}) =h_0 + \bm h \cdot \bm\sigma$, where 
\begin{align}
    h_x/|\bm h|&=2d_\mathrm{max}\sqrt{1-d_\mathrm{max}^2}\sin^2\theta_k,\nonumber \\
    h_y/|\bm h|&=2 d_\mathrm{max}\sin\theta_k\cos\theta_k,\nonumber \\
    h_z/|\bm h|&=(1-2 d_\mathrm{max}^2\sin^2\theta_k ), \label{eq:Ham}
\end{align}
where $\theta_k{=}\tan^{-1}(k_y/k_x)$ is the polar angle in momentum space and $|\bm h|{=}|\bm k|^2(m_+^{-1}{-}m_-^{-1})/4$ and $h_0{=}|\bm k|^2(m_+^{-1}{+}m_-^{-1})/4$ assuming $m_+^{-1}{>}m_-^{-1}$. 
The dispersion relations for the two bands are $\epsilon_\pm=|\bm k |^2/2m_\pm$.
Thus, $d_\mathrm{max}$ can be controlled without altering the band structure.
The pseudospin textures for the cases $d_\mathrm{max}{=}1$ and $d_\mathrm{max}{=}0$ are illustrated in Fig.~\ref{fig1}(b) and (c), respectively.

The foregoing analysis clearly indicates the anti-symmetric behavior in the transport scattering rate by pseudo-spin independent and dependent impurities with respect to $\gamma^2_0=|\bm \gamma|^2/3$, as shown in Fig.~\ref{fig2}(a,b). Let us thus focus on $U_{\text{pert}}= U_0 \sigma_0$, i.e., $|\bm \gamma|=0$.
In Figs.~2(c) and 2(d), the lower panels show the rotation of $\bm s_{\bm k}$ following the Fermi surface $\theta_k \in [0,2\pi]$ for $W=1$ (black) and $W=2$ with $d_{\rm{max}}=0, 0.5, 1$ {(blue, green, red).} The scattering probability $W_{\bm k \bm k'}$ plotted in Fig.~\ref{fig2}(c,d) oscillates $W$-times with an amplitude proportional to the $d^2_{\rm{max}}$. 
By integrating over states on the Fermi surface, the transport scattering rate is {evaluated as}
\begin{align}
    \biggl\langle \frac{1}{\tau^{\text{iso}}_{\bm k}} \biggr \rangle 
    =
    \frac{2\pi}{\hbar}\rho(\epsilon_F) \gamma_0^2 \left[ \frac{2-d_\mathrm{max}^2}{2} - \delta_{W1} \frac{d^2_{\rm{max}}}{4}\right], \label{eq:scat rate}
\end{align}
which highlights the significant role of $d_{\rm{max}}$ in the transport.

%
%

{Next, we consider the linear response function $L_{ij}$ for electric and thermal currents
\begin{align}
    \bpm j^c \\ j^{th} \epm = \bpm L_{11} & L_{12} \\ L_{21} & L_{22} \epm \bpm E \\ -\frac{\nabla T}{T} \epm.
\end{align}
We compute $L_{ij}$ and PF using $\tau_{\bm k}^{-1}$, Eqs.~\eqref{eq:distribution} and \eqref{eq:currents} (See SM for details).}
%
{We plot $L_{ij}$ and $PF$ as a function of the chemical potential $\mu$ at $d_\mathrm{max} =$ 0 (blue), 0.5 (green), and 0.9 (red) with $m_+ {=} -m_- {=} 1$ in Fig.~\ref{fig3}(a-d) by solid lines.}
%
%
The power factor is 
\begin{eqnarray}
PF(d_\mathrm{max})=\frac{2}{2-d_\mathrm{max}^2} PF(d_\mathrm{max}=0). \label{eq:PF}
\end{eqnarray}
This expression indicates that PF increases with $d_\mathrm{max}$, potentially doubling when $d_\mathrm{max}=1$ compared to the geometrically trivial case ($d_\mathrm{max}=0$), as shown in Fig.~\ref{fig3}(e).
This enhancement is attributed to the influence of quantum geometry on the scattering rate, highlighting that the manipulation of wavefunction geometry can significantly improve PF.
Note that the relation in Eq.~(\ref{eq:PF}) holds even when we consider the Coulomb-type impurity potential (See SM).

It is noteworthy that changes in the geometry of electronic states also influence the electronic thermal conductivity ($\kappa_{e}$). 
As a result, any increase in $PF$ due to changes in scattering time will also affect $\kappa_e$. However, the total thermal conductivity ($\kappa$) is not solely determined by electronic contributions. 
It also includes contributions from other factors such as phonons. The geometry of electronic wavefunctions does not alter these other contributions. 
Therefore, changing the quantum geometry of electric states to enhance $PF$ will ultimately lead to a better figure of merit ($ZT=\frac{PF}{\kappa} T$), enhancing the overall efficiency of the thermoelectric material.

\begin{figure}[t]
\includegraphics[width=85mm]{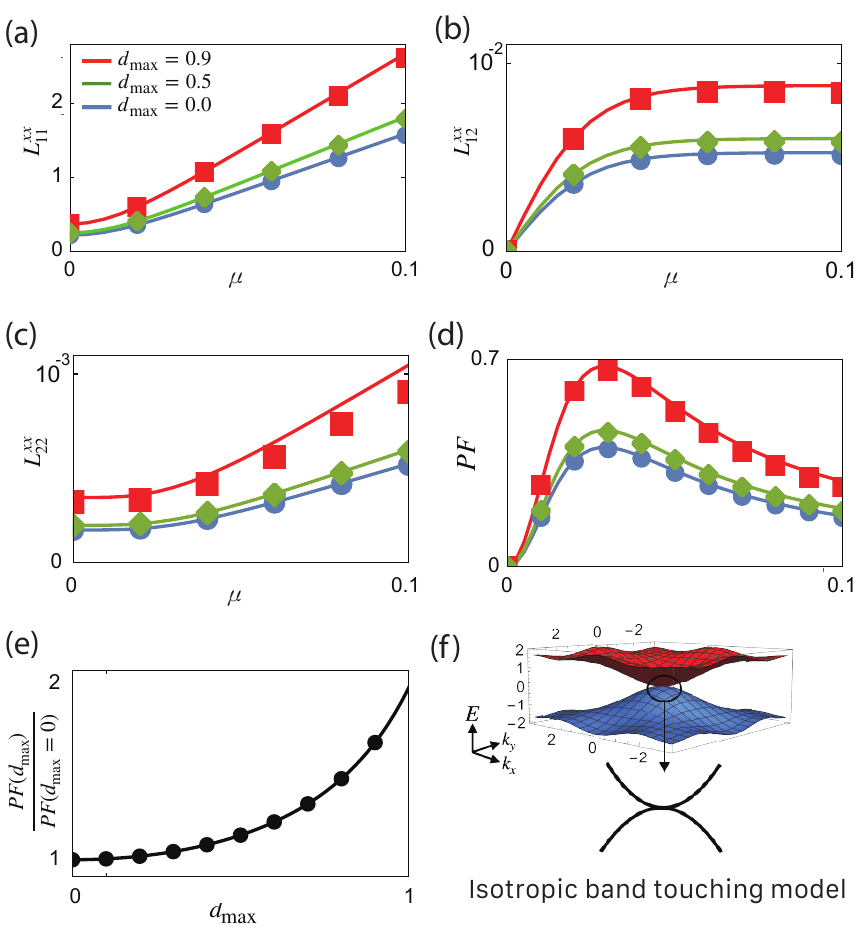} 
\caption{\label{fig3}
(a-d) Chemical potential $\mu$ dependence of (a) $L_{11}$, (b) $L_{12}=L_{21}$, (c) $L_{22}$ and (d) $PF$ for $d_\mathrm{max}=0$ (blue), 0.5 (green) and 0.9 (red). (e) represents the ratio between $PF(d_\mathrm{max})$ and $PF(0)$. The solid lines represent the results from the isotropic quadratic touching model in Eq.~\eqref{eq:Ham} with $m_+=-m_-=1$.
The dotted plots represent the results that are calculated from the lattice model Eq.~(\ref{eq:Ham_lat}). (f) Band structure of Eq.~\eqref{eq:Ham_lat} with $d_\mathrm{max}=0.9$.
In the calculation, we fix the parameters $T=0.01$, $k_B =\hbar=e=1$, $1/\tau(d_\mathrm{max}=0) = 0.01$, and $\gamma_0= const$. For the case of Coulomb-type impurity potential $\gamma_0$,
see SM.}
\end{figure}

To verify the validity of the results obtained from the continuum model, we investigate the following lattice model:
$\mathcal{H}_{\mathrm{lat}}(\bm{k}) = \sum_{\alpha=x,y,z } h_\alpha ({\bm{k}}) \sigma_\alpha$, where
\begin{align}
    h_x(\bm{k})&=d_\mathrm{max}\sqrt{1-d_\mathrm{max}^2}(1-\cos{k_y})),\nonumber \\
    h_y(\bm{k})&=d_\mathrm{max} \sin{k_x}\sin{k_y},\nonumber \\
    h_z(\bm{k})&=2-2d_\mathrm{max}^2-\cos{k_x}+(2 d_\mathrm{max}^2-1)\cos{k_y}. \label{eq:Ham_lat}
\end{align}
For $d_\mathrm{max}<1$, it has a band crossing point at $\Gamma$ point, as shown in Fig.~\ref{fig3}(f).
By taking the k-p expansion near the $\Gamma$ point, we derive the low energy effective model described in Eq.~(\ref{eq:Ham}) with $m^{-1}_+=1$ and $m^{-1}_-=-1$. 
As demonstrated in Fig.~\ref{fig3} {(square, diamond, and circular symbols)}, the response functions derived from the lattice model align well with those from the low energy effective model, with minor shifts due to lattice effects at high chemical potential $\mu$. 
This consistency clearly supports that the transport properties are primarily governed by $d_\mathrm{max}$, thereby validating the geometric transport formula based on $d_\mathrm{max}$.

\textit{Discussion.---}
We established a direct connection between the quantum distance and the scattering probability rate, which are essential for the charge or heat transport.
The maximum quantum distance $d_\mathrm{max}$ representing the intrinsic geometric property of a system has been shown to determine extrinsic thermoelectric transport coefficients averaged over an ensemble of disorder realizations. 
Our work shows that the entire distribution of quantum distances between states on the Fermi surface plays a crucial role in determining transport coefficients, thus significantly impacting thermoelectric performance.

Further investigations into various material systems, including experimental validations, are essential to fully explore the potential of geometric manipulation for enhancing thermoelectric efficiency.
Additionally, extending this work to multi-terminal thermoelectric devices and exploring extrinsic microscopic mechanisms for generating spin current from a quantum geometric perspective would be valuable directions for future research.

\begin{acknowledgements}
We thank Yusuke Kato and Haruki Watanabe for the useful discussions.
C-g. O is supported by Q-STEP, WINGS Program, the University of Tokyo. J.W.R is supported by the National Research Foundation of Korea (NRF) Grant
funded by the Korean government (MSIT) (Grant no.
2021R1A2C1010572 and 2021R1A5A1032996 and
2022M3H3A106307411) and the Ministry of Education(Grant no. RS-2023-00285390). K.W.K is supported by the National Research Foundation of Korea (NRF) grant funded by the Korean government(MSIT) (No.  2020R1A5A1016518)
\end{acknowledgements}

\bibliography{ref.bib}

\onecolumngrid

\clearpage
\newpage
\appendix



\section{Quantum distance between eigenstates on the Fermi surface}

When the crystal momentum changes $\bm k \rightarrow \bm{k+dk}$, an eigenstate changes by $d\ket \psi=\sum_\nu \ket{\partial_{k_\nu} \psi} d k_\nu$, and the distance between  the eigenstates is defined as $ds^2 = || \psi(\bm k)-\psi(\bm{k+d k})||^2 = \Braket {\sum_\mu \partial_{k_{\mu}}\psi \,d k_\mu | \sum_\nu \partial_{k_\nu} \psi \,d k_\nu}$. Using the quantum geometric tensor $Q_{\mu\nu}=\Braket {\partial_{k_{\mu}}\psi | \partial_{k_\nu} \psi}$, the distance is expressed as
\begin{align}
    ds^2 &= \sum_{\mu\nu} Q_{\mu\nu} dk_\mu dk_\nu, \\
    &= \sum_{\mu\nu} \frac 1 2 (Q_{\mu\nu}+Q_{\nu\mu}) dk_\mu dk_\nu, \\
    &\equiv \sum_{\mu\nu} g_{\mu\nu} dk_\mu dk_\nu,  
\end{align}
where quantum metric tensor $g_{\mu\nu}=g_{\nu\mu}$ is introduced, and in the second line $dk_\mu dk_\nu=dk_\nu dk_\mu$ is used. That is, the quantum metric tensor is the symmetric part of the quantum geometric tensor, $g_{\mu\nu}=\frac 1 2 (\Braket {\partial_{k_{\mu}}\psi | \partial_{k_\nu} \psi}+\Braket {\partial_{k_{\nu}}\psi | \partial_{k_\mu} \psi})=\Re \Braket {\partial_{k_{\mu}}\psi | \partial_{k_\nu} \psi}$. $Q_{\mu\nu}$ is in general a complex number and its imaginary part is associated with the accumulation of U(1) phase when a state makes a loop around an area $dk_xdk_y$, the Berry curvature $\Omega_{\mu\nu}=2\Im \Braket {\partial_{k_{\mu}}\psi | \partial_{k_\nu} \psi}$.
\newline
\newline
(add) the quantum metric is closely related to the quantum distance because it is the real part of the quantum geometric tensor, which is obtained as {$g_{ij}^n=\sum_{m(\neq n)}A_i^{nm}(\bm k)A_j^{mn}(\bm k)+c.c$ with the Berry connection $A_i^{nm}(\bm k)= i \braket{\partial_{k_i} u_n(\bm k)| u_m(\bm k)}$, where $u_n(\bm k)$ is the Bloch wave function for the $n$th band \cite{provost1980riemannian}.
\section{Isotropic quadratic band touching model}
In this section we show that the most general Hamiltonian describing the isotropic quadratic band touching can be written with three parameters (the mass of upper/lower band $m_\pm$ and the maximum quantum distance $d_{\mathrm{max}}$) and the eigen states and pseudo spin vector can be expressed only in terms of $d_{\mathrm{max}}$ and $\theta_{\bm k}$. 
\newline
\newline
The most general 2D quadratic band touching Hamiltonian \cite{rhim2019classification} is
    \begin{eqnarray}
        \mathcal{H}({\bm{k}}) = \sum_{\alpha=0,x,y,z } f_\alpha ({\bm{k}}) \sigma_\alpha ,
    \end{eqnarray}
where $h_0(\bm k) = b_1 k_x^2 +b_2 k_x k_y + b_3 k_y^2$ and
    \begin{align}
    &&h_x(\bm k) =t_6 k_y^2, 
    &&h_y(\bm k) =t_4 k_x k_y+ t_5 k_y^2, 
    &&h_z(\bm k) = t_1 k_x^2 + t_2 k_x k_y +t_3 k_y^2,
    \end{align}
The condition for the isotropic band touching model is
\begin{align}
 h_0^2 =|\bm k|^4\frac 1 {16} \left(\frac{1}{m_+}+\frac{1}{m_-} \right)^2, \,\,\,\,\,\, h_x^2+h_y^2+h_z^2=|\bm k|^4\frac 1 {16} \left(\frac{1}{m_+}-\frac{1}{m_-} \right)^2, 
\end{align}
such that the momentum-energy dispersion relation is $\epsilon_\pm=|\bm k|^2/2m_\pm$. The straightforward algebra provides that 
\begin{align}
    &t_1=\frac{1}{4} \left( \frac{1}{m_+} - \frac{1}{m_-} \right),  &t_2=0, \,\,\,\,\,\,\,\, &t_3 = \frac{1-2d^2_{\mathrm{max}}}{4} \left( \frac{1}{m_+} - \frac{1}{m_-} \right),  &t_4 = \frac{d_{\mathrm{max}}}{2} \left( \frac{1}{m_+} - \frac{1}{m_-} \right), \,\,\,\,\,\,\,\, &t_5=0, \nonumber \\ &b_1 =\frac{1}{4}\left( \frac{1}{m_+} + \frac{1}{m_-}\right),   &b_2=0,  \,\,\,\,\,\,\,\, &b_3=\frac{1}{4}\left( \frac{1}{m_+} + \frac{1}{m_-}\right), &  & 
\end{align}
where note that parameter $d_{\mathrm{max}}$ is involved in the Hamiltonian while its eigenvalues are only determined by mass $m_\pm$ and momentum. Its physical meaning as the maximum quantum distance will be clear in the following. Eigenvectors at $\bm{k}=(k \cos{\theta_{\bm k}},k \sin{\theta_{\bm k}})$ are
\begin{align}
    \ket{+,\bm{k}} &= \bpm \sqrt{1-d_\mathrm{max}^2} \sin\theta_{\bm k} - i \cos\theta_{\bm k} \\ d_\mathrm{max} \sin\theta_{\bm k} \epm,\\
    \ket{-,\bm{k}} &= \bpm -d_\mathrm{max} \sin\theta_{\bm k}\\ \sqrt{1-d_\mathrm{max}^2} \sin\theta_{\bm k} + i \cos\theta_{\bm k}  \epm,
\end{align}
which is independent of $m_\pm$. Using the eigenvector of the conduction band, we can compute the pseudospin vector, $\bm s=\bra{+\bm k} \hat S \ket{+\bm k}$, 
\begin{align}
    \bm s = (d_{\mathrm{max}}\sqrt{1-d_{\mathrm{max}}^2} (1-\cos2\theta_{\bm k}), -d_{\mathrm{max}}\sin 2\theta_{\bm k}, 1-d_{\mathrm{max}}^2+d_{\mathrm{max}}^2\cos2\theta_{\bm k}), 
\end{align}
which is normalized $|\bm s|=1$, and the vector makes a circle around $\bm s_0=(d_{\mathrm{max}}\sqrt{1-d_{\mathrm{max}}^2}, 0, 1-d_{\mathrm{max}}^2)$ with radius $d_{\mathrm{max}}^2$. Note that the eigenstate can be generated by the SU(2) rotation around $\bm s_0$ axis: 
\begin{align}
     \ket{+, \bm k} = e^{i\theta_k (\hat s_0 \cdot \bm \sigma)} \bpm 1 \\ 0 \epm ,      \,\,\,\,\,\,\, \ket{-, \bm k} = e^{i(\frac{\pi}{2}+\theta_k) (\hat s_0 \cdot \bm \sigma)} \bpm 1 \\ 0 \epm,
\end{align}
where $\hat s_0 = (d_{\mathrm{max}}, 0, \sqrt{1-d_{\mathrm{max}}^2})$. Because the pseudospin vector also rotates around $\hat s_0$, the Hamiltonian has the rotation symmetry by $\hat U_\theta$ mentioned in the main text with $W=1$ for the above model.

\section{Boltzmann transport}
When a system reached a steady state under an electric field and temperature gradient, the distribution of electron in n at momentum k stays constant in  time with scatterings back to the Fermi Dirac distribution:  
\begin{align}
    \frac {df} {dt} &= \frac{\partial f}{\partial t} + \frac{\partial f}{\partial r_i} \dot r_i + \frac{\partial f}{\partial k_i} \dot k_i =  \frac{\partial f}{\partial \xi} \frac{\partial \xi}{\partial r_i} \dot r_i +\frac{\partial f}{\partial \xi} \frac{\partial \xi}{\partial k_i} \dot k_i , 
\end{align}
where $\xi = (\epsilon_{nk}-\mu)/k_BT$ is introduced, and then $\frac{\partial \xi}{\partial r_i}=-\partial_{r_i} (\mu/T)$, $\frac{\partial \xi}{\partial k_i} = \frac{1}{k_B T} \partial_{k_i} \epsilon_{nk}$. The velocity in real space and momentum space is  $\dot r_i = \frac 1 {\hbar} \partial_{k_i} \epsilon_{nk}$, $\dot k_i = e E_i /\hbar$, respectively. 
\begin{align}
    \frac{df}{dt} =   \frac{\partial f}{\partial \xi} \left[ \partial_{r_i} \xi +  \frac{1}{k_B T} {e E_i}\right] \frac 1 {\hbar} \partial_{k_i} \epsilon_{nk} = - \frac{f(n,k,r)-f_0}{\tau_{nk}} , 
\end{align}
from which one can obtain the expression \eqref{eq:distribution}. 
The linear response functions of transport can be extracted 
\begin{align}
    \bpm j^c \\ j^{th} \epm = \bpm L_{11} & L_{12} \\ L_{21} & L_{22} \epm \bpm E \\ -\frac{\nabla T}{T} \epm.
\end{align}
combined with the electric current and energy current density in \eqref{eq:currents}, the coefficient are explicitly 
\begin{align}
    L^{ij}_{\mu\nu} =-\frac{e^4}{V}\left(\frac{k_BT}{e}\right)^{\mu+\nu}\sum_{n,\bm k} \frac{(\hbar/\tau_{n\bm{k}})^{-1}}{2+2\cosh\xi} \frac{(\partial_{k_i}\xi) (\partial_{k_j}\xi)}{\xi^{2-\mu-\nu}} ,\label{eq:res_func_1}
\end{align}
where $\mu,\nu=1,2$ is for electrical and heat conductivity, respectively.
Note that $L_{12}=L_{21}$ for time-reversal symmetric materials according to the Onsager's reciprocal theorem~\cite{onsager1931reciprocal}.
$PF$ can be described using $L_{ij}$ as 
\begin{align}
PF=\frac{1}{T^2}\frac{L_{12}^2}{L_{11}}
\end{align}

\section{Disorder averaged transport scattering rates}
For the most general type of scattering potential $U_\text{pert}=\sum_{i=0,x,y,z} U_i(\bm{r})\sigma_i$ the scattering probability under the Born approximation is  
\begin{align}
    W_{\bm{k'k}} &=  \frac{2\pi}{\hbar}\Bigl[ (\bm v_{\bm q}\cdot \bm v_{\text{-}\bm q}- |v_{\bm q,0}|^2)\frac{1-\bm s_{\bm k} \cdot \bm s_{\bm{k'}}}{2} + \sum_{i,j=0,x,y,z} (v_{\bm q,i}\bm s_{\bm k,i}) (v_{\text{-}\bm q,j}\bm s_{\bm k',j}) \Bigr] \delta(\epsilon_{\bm{k'}}-\epsilon_{\bm k}), \label{seq:scat_prob}
\end{align}
where the scattering potential in the momentum space $v_{\bm q,i}=\frac{1}{L^2}\int d^2\bm{r}\,U_i(\bm{r})e^{-i\bm q\cdot\bm{r}}$, the momentum transfer by a scattering $\bm q=\bm k'-\bm k$, $s_{\bm k,i}\equiv (s_{\bm k})_i$ and $(s_{\bm k})_{i=0}=1$ are introduced. 

The disorder averaging: 
\begin{align}
    \langle \cdots \rangle = \int \prod_i dU_i e^{-\frac{1}{2\gamma_i^2}\sum_i \int \frac{d^2\bm r}{V} U^2_i(\bm r)} (\cdots),
\end{align}
where index $i=0,x,y,z$ for non-magnetic and magnetic scattering potentials. For example, 
\begin{align}
    \langle U_i(\bm r) U_j(\bm r') \rangle = V\gamma_j^2 \delta(\bm r-\bm r') \delta_{ij}
\end{align}
which is for the delta function correlated impurity scattering potential. (This can be extended to Coulomb scattering potential). The correlation between the Fourier transformed scattering potential, 
\begin{align}
    \langle v_i(\bm q) v_j(\bm q') \rangle = \gamma_j^2 \delta(\bm q+\bm q') \delta_{ij}, 
\end{align}
which is used to remove the crossing terms. As a result, 
\begin{align}
    \langle W_{\bm{k'k}}\rangle &=  \frac{2\pi}{\hbar}\Bigl[ \langle \bm v_{\bm q}\cdot \bm v_{\text{-}\bm q}\rangle\frac{1-\bm s_{\bm k} \cdot \bm s_{\bm{k'}}}{2}  +\langle|v_{\bm q,0}|^2\rangle \frac{1+\bm s_{\bm k} \cdot \bm s_{\bm{k'}}}{2} +\sum_{i=x,y,z}\langle |v_{\bm q,i}|^2\rangle  \bm s_{\bm k,i}\bm s_{\bm k',i}\Bigr] \delta (\epsilon_k - \epsilon_{k'}), \\
    &=  \frac{2\pi}{\hbar}\Bigl[ |\bm\gamma|^2\frac{1-\bm s_{\bm k} \cdot \bm s_{\bm{k'}}}{2}  +\gamma_0^2 \frac{1+\bm s_{\bm k} \cdot \bm s_{\bm{k'}}}{2} +\sum_{i=x,y,z}\frac{|\bm\gamma|^2}{3} \bm s_{\bm k,i}\bm s_{\bm k',i}\Bigr] \delta (\epsilon_k - \epsilon_{k'}), \\
    &=  \frac{2\pi}{\hbar}\Bigl[ \frac{|\bm\gamma|^2}{3}(1+d_{\bm k \bm k'}^2)  +\gamma_0^2 (1-d_{\bm k \bm k'}^2) \Bigr] \delta (\epsilon_k - \epsilon_{k'}),
\end{align}

In the second line $\langle |v_{\bm q,i}|^2\rangle=|\bm \gamma^2|/3$ is used, and in the third line the quantum distance is introduced $\bm s_{\bm k}\cdot \bm s_{\bm k'}=1-2d_{\bm k \bm k'}^2$. For an isotropic system with extra spin rotation symmetry, the pseudo spin vector is $\bm s_{\bm k }=\bm s_{0} + \bm s_{\perp\alpha} \cos(W\theta_{\bm k}) + \bm s_{\perp\beta}\sin(W\theta_{\bm k})$, where $\bm s_{0,\perp\alpha,\perp\beta}$ are orthogonal each other,$|\bm s_0|\text{=}\sqrt{1-d_\mathrm{max}^2}$, and $|\bm s_{\perp \alpha}|\text{=}|\bm s_{\perp \beta}|\text{=}d_\mathrm{max}$. The momentum dependent quantum distance is
\begin{align}
    d^2_{\bm{kk'}} = d_\mathrm{max}^2\frac{1-\cos W\theta_{kk'}}{2}
\end{align}
where where $d_\mathrm{max}=\text{max}_{\bm k'\in \text{FS}}\left[d_{\bm{kk'}}\right]$ the maximum quantum distance on the Fermi surface, $W$ is an integer. 
\begin{align}
    \biggl\langle \frac{1}{\tau_k^{\text{iso}}} \biggr\rangle 
    &=  \sum_{k'\in \text{FS}}  \langle W_{kk'} \rangle (1-\cos\theta_{kk'}),\\
    &=\frac{2\pi}{\hbar}\frac{\rho(\epsilon_F)}{2} \left[\gamma_0^2 (2-d_\mathrm{max}^2) + 
    \frac{|\bm \gamma|^2}{3} (2+d_\mathrm{max}^2) \right]+d_\mathrm{max}^2\frac{2\pi}{\hbar}\rho(\epsilon_F)\frac{|\bm \gamma|^2-3\gamma_0^2}{12}\delta_{W1}.
\end{align}

\section{Higher order corrections to the Fermi Golden rule}
The scattering probability involves the following quantity:
\begin{align}
    w^{}_{kk'} &= |\bra {k'} \left( v_{q,i}\sigma_i \right) \ket k|^2 
    = \bra {k'} \left(v_{q,i}\sigma_i \right)\ket k \bra k \left( v_{-q,m}\sigma_m \right)\ket {k'} , \\
    &=\bra {k'} \left(v_{q,i}\sigma_i\right)\frac 1 2 \left( s_{k,j}\sigma_j\right) \left( v_{-q,m}\sigma_m \right) \ket {k'} ,\\
    &=\frac 1 2 \left(v_{q,i} v_{-q,m}\right)\left( s_{k,j} s_{k',l} \right), 
\end{align}
where the summation over repeated indices ($i,j,m=\{0,x,y,z\}$) are omitted. $v_{q,0}=1$ and $s_{k,0}=1$. $s_{k',l}=\bra{k'} \sigma_i \sigma_j \sigma_m \ket{k'}$. The scattering probability expressed in terms of pseudospin vectors. After the disorder averaging, $\langle v_{q,i} v_{-q,m}\rangle =\gamma_i^2 \delta_{im}$. Then, if $j=i$, $s_{k',l}=s_{k',j}$, if $j\neq i$, $s_{k',l}=-s_{k',j}$. When $\gamma_x^2=\gamma_y^2=\gamma_z^2=|\bm \gamma|^2/3$, the above can be reduced to 
{
\begin{eqnarray}
w^{}_{kk'} = 
\Bigl[ \frac{|\bm\gamma|^2}{3}(1+d_{\bm k \bm k'}^2)  +\gamma_0^2 (1-d_{\bm k \bm k'}^2) \Bigr]
\end{eqnarray}
}
The next order is 
\begin{align}
    w^{(2)}_{kk''k'} &= |\bra {k'} \left( v_{q,i}\sigma_i \right) \ket{k''}\bra{k''} \left( v_{q',l}\sigma_l \right) \ket k|^2 , \\
    &= \bra {k'} \left( v_{q,i}\sigma_i \right) \ket{k''}\bra{k''} \left( v_{q',l}\sigma_l \right) \ket k  \bra{k} \left( v_{-q',l'}\sigma_{l'} \right) \ket {k''}\bra{k''} \left( v_{-q,i'}\sigma_{i'} \right) \ket{k'} , \\
    &= \bra {k'} \left( v_{q,i}\sigma_i \right) \frac 1 2 \left( s_{k'',j}\sigma_j\right) \left( v_{q',l}\sigma_l \right) \frac 1 2 \left( s_{k,j'}\sigma_{j'}\right) \left( v_{-q',l'}\sigma_{l'} \right)\frac 1 2 \left( s_{k'',j''}\sigma_{j''}\right) \left( v_{-q,i'}\sigma_{i'} \right) \ket{k'} , \\
    &= \frac{1}{8}  \left(v_{q,i} v_{q',l} v_{-q',l'} v_{-q,i'}  \right) \left( s_{k'',j} s_{k,j'}s_{k'',j''} s_{k',j'''}\right)
\end{align}
 where $s_{k',j'''}=\bra{k'} \sigma_i \sigma_j \sigma_l \sigma_{j'} \sigma_{l'} \sigma_{j''} \sigma_{i'} \ket{k'}$, {$\bm{q}=\bm{k'-k''}$, and $\bm{q'}=\bm{k''-k}$}. They are all expressed in terms of the multiplication of pseudospin vectors. After the disorder averaging, the scattering probability becomes isotropic in spinor space, then the scattering probability can be expressed in terms of the quantum distance $d^2_{kk''}$ and $d^2_{k'k''}$. 
%

After taking the disorder averaging, we get
\begin{align}
    w^{(2)}_{kk''k'} 
    &=  \braket{v_{q,i}v_{q',l}v_{-q',l'}v_{-q,i'}}
    \bra {k'} \sigma_i \ket{k''}\bra{k''}  \sigma_l  \ket k  \bra{k}  \sigma_{l'} \ket {k''}\bra{k''}  \sigma_{i'}  \ket{k'},\\
\end{align}
Here, $\braket{v_{q,i}v_{q',l}v_{-q',l'}v_{-q,i'}}=\Big(\braket{v_{q,i}v_{q',l}}\braket{v_{-q',l'}v_{-q,i'}}+\braket{v_{q,i}v_{-q,i'}}\braket{v_{q',l}v_{-q',l'}}+\braket{v_{q,i}v_{-q',l'}}\braket{v_{q',l}v_{-q,i'}}
    \Big)$, and we consider an approximation $\braket{v_{q,i}v_{q',l}v_{-q',l'}v_{-q,i'}} \approx \braket{v_{q,i}v_{-q,i'}}\braket{v_{q',l}v_{-q',l'}}$, since 
$\braket{v_{q,i}v_{q',l}}\propto \delta(\bm{k'}-\bm{k})$ and $ \braket{v_{q,i}v_{-q',l'}} \propto \delta(\bm{k'}+\bm{k}-2\bm{k''})$, and the related terms only give small deviations. Hence,
\begin{align}
    w^{(2)}_{kk''k'} 
    &=  \braket{v_{q,i}v_{-q,i'}}\braket{v_{q',l}v_{-q',l'}}
    \bra {k'} \sigma_i \ket{k''}\bra{k''}  \sigma_l  \ket k  \bra{k}  \sigma_{l'} \ket {k''}\bra{k''}  \sigma_{i'}  \ket{k'},\\
    &= \braket{v_{q,i}v_{-q,i'}}\bra {k'} \sigma_i \ket{k''}\bra{k''}  \sigma_{i'}  \ket{k'}
    \braket{v_{q',l}v_{-q',l'}}
     \bra{k''} \sigma_l  \ket k  \bra{k}  \sigma_{l'} \ket {k''},\\
    &= \Bigl[ \frac{|\bm\gamma|^2}{3}(1+d_{\bm k' \bm k''}^2)  +\gamma_0^2 (1-d_{\bm k' \bm k''}^2) \Bigr] \Bigl[ \frac{|\bm\gamma|^2}{3}(1+d_{\bm k \bm k''}^2)  +\gamma_0^2 (1-d_{\bm k \bm k''}^2) \Bigr] \label{eq:C15}.
\end{align}

That is, $w^{(2)}_{kk''k'}\simeq w^{}_{kk''}w^{}_{k''k'}$.

The Fermi Golden rule including the higher order corrections~\cite{wang2023generalized} is
\begin{align}
    W_{kk'} &= \frac{2\pi}{\hbar} |\bra {k'}  U_{\text{pert}} \left(I - G^{\text{ret}}_0 U_{\text{pert}} \right)^{-1}  \ket k|^2 \delta (\epsilon_k - \epsilon_{k'}), \\
    &= \frac{2\pi}{\hbar} \delta (\epsilon_k - \epsilon_{k'})\sum_{n=1,2,\cdots}W^{(n)}_{kk'},  \label{eq:C17}
\end{align}
where $G^{\text{ret}}_0 = \sum_{k''}\ket{k''}\bra{k''}/(\epsilon_k-\epsilon_{k''}+i0^+)$, and 
\begin{align}
W^{(1)}_{kk'}&=w^{}_{kk'},\\
    W^{(2)}_{kk'}&=\sum_{k''} \frac{ w^{(2)}_{kk''k'} }{(\epsilon_k-\epsilon_{k''}+i0^+)(\epsilon_k-\epsilon_{k''}-i0^+)}. 
\end{align}
After disorder averaging, the transport scattering rate $\langle 1/\tau_{k}\rangle$ will be in terms of $\sum_{k'\in \text{FS}} d^2_{kk'}$, $\sum_{k',k''\in \text{FS}} d^2_{kk''}d^2_{k''k'}$, etc. As a result, the distribution of quantum distance on the Fermi surface (including higher moments) will be reflected in the rate. 
\newline

\noindent \textit{Dressed Green function}: Consider the following quantity. 
\begin{align}
\bra {k'}  T_+ \ket k \bra {k}  T_- \ket {k'}=
    \bra {k'}  U_{\text{pert}} \left(I - G^{r}_0 U_{\text{pert}} \right)^{-1}  \ket k \bra {k}  U_{\text{pert}} \left(I - G^{a}_0 U_{\text{pert}} \right)^{-1}  \ket {k'}.
\end{align}
When one performs the disorder averaging, if the pairing of two scattering potentials is selected in the first (second) T-matrix $T_+$ ($T_-$), it renormalizes the retarded (advanced) Green function: 
\begin{align}
    G^{r} 
    &=  (\epsilon -H_0 - \Sigma^r)^{-1},  \\
    &= G^r_0 + G^r_0 \Sigma^r G^r_0 + G^r_0 \Sigma^r G^r_0 \Sigma^r G^r_0 + \cdots,
\end{align}
where $G^r_0 = \sum_k \frac{\ket k \bra k}{\epsilon- \epsilon_k + i0^+}$. 
where the self energy can be obtained by collecting non-crossing diagrams (figure)
\begin{align}
    \Sigma^r &= \sum_{i=0,x,y,z}(v_{q,i}\sigma_i) G^r_0 (v_{-q,i}\sigma_i), \\
    &= \sum_{i=0,x,y,z}|v_{q,i}|^2 \sum_k \frac{\sigma_i \ket k \bra k \sigma_i }{\epsilon- \epsilon_k + i0^+}, \\
    &= \sum_{i,j=0,x,y,z} |v_{q,i}|^2 \sum_k \frac{\sigma_i \frac 1 2 s_{k,j}\sigma_j \sigma_i }{\epsilon- \epsilon_k + i0^+}, 
\end{align}
where $\ket k \bra k = \sum_{j=0,x,y,z}\frac 1 2 s_{k,j} \sigma_j$ and $s_{k,0}=1$. The summation over momentum $k$ can be split to the integration over angle $\theta_k$ and the magnitude $|k|$. Then, the latter is converted to  energy integration. We focus on the imaginary part of the self energy: 
\begin{align}
    \text{Im} [\Sigma^r] =-\pi \rho(\epsilon_F) \sum_{i,j=0,x,y,z}  \sigma_i \sigma_j \sigma_i |v_{q,i}|^2 \int \frac{d\theta_k}{2\pi} \frac 1 2 s_{\theta_k,j},  
\end{align}
which contains a typical pseudospin independent component ($\sim \sigma_0$) and dependent component ($\sim \sigma_{\hat n}$), where $\hat n$ is the averaged pseudospin direction on the Fermi surface. 
Explicitly, 
\begin{align}
    \text{Im} [\Sigma^r] &=-\frac{\pi} 2 \rho(\epsilon_F) \left[ \sum_{i=0,x,y,z} |v_{q,i}|^2\sigma_0 + \sum_{j=x,y,z} \left( |v_{q,0}|^2-|\bm v_{q}|^2 +2|v_{q,j}|^2  \right)  \sigma_j\int \frac{d\theta_k}{2\pi}  s_{\theta_k,j}   \right] , \\
    &=-\frac{\pi} 2 \rho(\epsilon_F) \left[\gamma_0^2 (\sigma_0+\sum_{j=x,y,z}\sigma_j \int \frac{d\theta_k}{2\pi}  s_{\theta_k,j} ) + \frac 1 3 |
    \bm \gamma|^2 \sum_{j=x,y,z}(\sigma_0- \sigma_j \int \frac{d\theta_k}{2\pi}  s_{\theta_k,j} )  \right]
\end{align}
where in the second line $ |v_{q,0}|^2{=}\gamma_0^2$ and $ |v_{q,x}|^2{=}|v_{q,y}|^2{=}|v_{q,z}|^2{=}\frac 1 3 |\bm \gamma|^2$ are used. The pseudospin dependent scattering rate is interesting that we leave for the future study. In the following, we consider a situation with $\int d\theta_k s_{\theta_k,j}=0$ and therefore $\text{Im} [\Sigma^r] =-\frac{\pi} 2 \rho(\epsilon_F) (\gamma_0^2 + |\bm \gamma|^2)\sigma_0$. This is the case when pseudospin $\bm s$ make a circle with $\bm s_0=0$. In this case, even though $d_{\text{max}}=1$, we symbolically keep the notation. The computation for $d_{\text{max}}<1$ will require more careful treatment of pseudospin structure of $\Sigma^r$. 
\newline

\noindent \textit{The vertex correction}: Let us consider the pairing of two scattering potentials where one is selected from $T_+$ and the other from $T_-$. Using the disorder averaged Green's function $G^{r,a}$, (add a figure of a diagram where two curvy line connecting retarded and advanced GF lines in parallel)
\begin{align}
    W^{(2)}_{kk'}&=\sum_{k''} \frac{ w^{(2)}_{kk''k'} }{(\epsilon_k-\epsilon_{k''}+i\eta)(\epsilon_k-\epsilon_{k''}-i\eta)}, \\
    &=\int \frac{d\theta_{k''}}{2\pi} w^{(2)}_{\theta_k \theta_{k''}\theta{k'}}  \int d\epsilon_{k''}\rho(\epsilon_{k''})\frac{1 }{(\epsilon_k-\epsilon_{k''}+i\eta)(\epsilon_k-\epsilon_{k''}-i\eta)}, \\
    &=\int \frac{d\theta_{k''}}{2\pi} w^{(2)}_{\theta_k \theta_{k''}\theta{k'}}  \int d\epsilon_{k''}\rho(\epsilon_{k''})\left[ \frac{1 }{\epsilon_k-\epsilon_{k''}-i\eta} - \frac{1 }{\epsilon_k-\epsilon_{k''}+i\eta} \right]\frac{1}{2i \eta}, \\
    &=\frac{2}{\gamma_0^2 + |\bm \gamma|^2} \int \frac{d\theta_{k''}}{2\pi} w^{(2)}_{\theta_k \theta_{k''}\theta_{k'}}, 
\end{align}
where in the second line we consider $w^{(2)}_{k k''k'}=w^{(2)}_{\theta_k \theta_{k''}\theta{k'}}$ is only a function of angles of momentum. In the third line, $\eta = -\text{Tr Im}\Sigma^r= {\pi}\rho(\epsilon_F) (\gamma_0^2 + |\bm \gamma|^2)$. The above calculation indicates that $W^{(1)}_{kk'}$ and $W^{(2)}_{kk'}$ share the same scaling on the disorder strength, $\sim \gamma_j^2$. 
\begin{align}
     W^{(2)}_{kk'}
     &\simeq \frac{1}{\gamma_0^2 + |\bm \gamma|^2} \int \frac{d\theta_{k''}}{2\pi} w^{}_{kk''}w^{}_{k''k'}, 
\end{align}
One can continue to take into account higher order processes, (add a figure with three curly lines connecting GR and GA lines, then n-lines. ladder diagram). 
\begin{align}
     W^{(3)}_{kk'}
     &\simeq \left(\frac{1}{\gamma_0^2 + |\bm \gamma|^2} \right)^2 \int \frac{d\theta_{k_2}}{2\pi} \int \frac{d\theta_{k_3}}{2\pi}  w^{}_{k k_2}w^{}_{k_2k_3}w^{}_{k_3k}, \\
     W^{(n)}_{kk'}&\simeq \left(\frac{1}{\gamma_0^2 + |\bm \gamma|^2} \right)^{n-1} \int  \prod_{j=2}^n\left( \frac{ d\theta_{k_j}}{2\pi}\right)   w^{}_{k k_2}w^{}_{k_2k_3}\cdots w^{}_{k_jk_{j+1}}\cdots w^{}_{k_{n}k}, 
\end{align}
which include $2n$ power of quantum distances. The disorder averaged generalized Fermi Golden rule is the summation of them all \eqref{eq:C17}. As a result, including all powers of disorder scatterings, the distribution of quantum distances on the Fermi surface determines the transition probability. That is, the information of whole moments of quantum distances is involved in the transition probability 
\newline

\noindent 
\textit{Example: Isotropic model:--} When a system has the orbital-spin rotation symmetry by $\hat U_{\theta_k,W}$, the higher order multiplication of quantum distances can be lowered down to the quadratic:
{\begin{eqnarray}
    \int \frac{d\theta_{k''}}{2\pi} \left(1-2\frac{d^2_{kk''}}{d_{\text{max}}^2} \right) \left(1-2\frac{d^2_{k''k'}}{d_{\text{max}}^2} \right) 
    = \int \frac{d\theta_{k''}}{2\pi} \cos W\theta_{kk''} \cos W\theta_{k''k'}
    = \frac 1 2 \left(1-2\frac{d^2_{kk'}}{d_{\text{max}}^2} \right) 
    \end{eqnarray}
}    
Try to reduce the following:
{
\begin{align}
    w_{kk'} &=  \gamma_s^2\frac{2+d^2_{\text{max}}}{2}+\gamma_0^2 \frac{2-d^2_{\text{max}}}{2} + \frac{\gamma_0^2-\gamma_s^2}{2}d^2_{\text{max}}\cos W\theta_{kk'},\\
    \int \frac{d\theta_{k_2}}{2\pi} w^{}_{kk_2}w^{}_{k_2k'} &= 
     \left[ \gamma_s^2\frac{2+d^2_{\text{max}}}{2}+\gamma_0^2 \frac{2-d^2_{\text{max}}}{2}\right]^2 + \left[\frac{\gamma_0^2-\gamma_s^2}{2}d^2_{\text{max}} \right]^2 \frac 1 2 \cos W\theta_{kk'},\\
     \int  \prod_{j=2}^n\left( \frac{ d\theta_{k_j}}{2\pi}\right)   w^{}_{k k_2}\cdots \cdots w^{}_{k_{n}k} &= \left[ \gamma_s^2\frac{2+d^2_{\text{max}}}{2}+\gamma_0^2 \frac{2-d^2_{\text{max}}}{2}\right]^n + \left[\frac{\gamma_0^2-\gamma_s^2}{2}d^2_{\text{max}} \right]^n \frac 1 {2^{n-1}} \cos W\theta_{kk'},
\end{align}
}
where $\gamma_s^2=|\bm \gamma|^2/3$ is introduced. As a result, 
\begin{align}
    W_{kk'} 
    &= \frac{2\pi}{\hbar} \delta (\epsilon_k - \epsilon_{k'})\sum_{n=1,2,\cdots}W^{(n)}_{kk'},\\
    &=\frac{2\pi}{\hbar} \delta (\epsilon_k - \epsilon_{k'})\sum_{n=1,2,\cdots}
    \left(\frac{1}{\gamma_0^2 + 3 \gamma_s^2} \right)^{n-1} \left\{ \left[ \gamma_s^2\frac{2+d^2_{\text{max}}}{2}+\gamma_0^2 \frac{2-d^2_{\text{max}}}{2}\right]^n + \left[\frac{ \gamma_0^2-\gamma_s^2}{2}d^2_{\text{max}} \right]^n \frac 1 {2^{n-1}} \cos W\theta_{kk'} \right\}, 
\end{align}
which is the scattering probability including all orders of disorder scatterings (taking non-crossing diagrams). We can next obtain the transport scattering from \eqref{eq:scat rate}. 
The summation can be performed and note that we assumed $d_{\text{max}}=1$ which makes $\text{Im} [\Sigma^r] =-\frac{\pi} 2 \rho(\epsilon_F) (\gamma_0^2 + |\bm \gamma|^2)\sigma_0$: {
\begin{align}
   \frac{1}{\tau^{(sc)}_k} &= \sum_{k'} W_{kk'} 
    =\frac{2\pi}{\hbar} \rho(\epsilon_F) \left( \gamma_s^2\frac{2+d^2_{\text{max}}}{2}+\gamma_0^2 \frac{2-d^2_{\text{max}}}{2}\right) 
    \left[1-\frac{1}{\gamma_0^2+3\gamma_s^2}(\gamma_s^2\frac{2+d^2_{\text{max}}}{2}+\gamma_0^2 \frac{2-d^2_{\text{max}}}{2})\right]^{-1} , \\
    &\frac{1}{\tau^{(tr)}_k} = \frac{1}{\tau^{(sc)}_k}-\frac{2\pi}{\hbar} \rho(\epsilon_F)\frac{d_\mathrm{max}^2(\gamma_0^2-\gamma_s^2)}{4+d_\mathrm{max}^2(\gamma_0^2-\gamma_s^2)}\delta_{W1},
\end{align}
}

\section{Coulomb-type impurity potential in isotropic quadratic band touching model}
In the case of a Coulomb-type impurity potential
, $\gamma_0 = v_{\bm{q},0}$ is a function of momentum $\bm{q}$ . The Fourier transformed potential becomes $v_{\bm{q},0}=2\pi e^2/\epsilon_0( |\bm{q}|+q_s)\sigma_0$, and $W_{kk'}$ becomes as follows:
\begin{align}
    W_{\bm{k'}{\bm{k}}}^n = n_\mathrm{imp}\frac{2\pi}{\hbar}\tilde{v}_{\text{imp}}(\bm{k-k'})^2 \left[1-d_{\text{HS},n}^2({\bm{k}},\bm{k'})\right]\delta(\epsilon_{n,\bm{k'}}-\epsilon_{n,{\bm{k}}}), 
\end{align}
where $n_\mathrm{imp}$ is the density of impurities.
The static dielectric function is written as 
\begin{eqnarray}
 \epsilon(\bm{q})=1+ \frac{2\pi e^2}{\epsilon_0 q}\Pi(q),
\end{eqnarray}
where $\Pi(q)$ is the polarization function.
In the long-wavelength limit, the dielectric function is written as 
\begin{eqnarray}
 \epsilon(q)=1+ \frac{2\pi e^2}{\epsilon_0 q}\Pi(0),
\end{eqnarray}
The screening constant $q_s$ is written as 
\begin{eqnarray}
 q_s= \frac{2\pi e^2}{\epsilon_0 }\Pi(0).
\end{eqnarray}
From the compressibility sum-rule, 
\begin{eqnarray}
\Pi(0)=\int^\infty_{-\infty} (-\frac{\partial f^{(0)}(\epsilon)}{\partial \epsilon})\rho(\epsilon)d\epsilon,
\end{eqnarray}
where $\rho(\epsilon)$ is the density of states. Therefore, $q_s=2\pi e^2 \rho(\epsilon_F)/\epsilon_0=2me^2/\epsilon$, where $m$ is the effective mass.

If we consider the quadratic isotropic band touching model, we get 
\begin{align}
    \frac{1}{\tau_{\bm{k}_F}} =
    n_\mathrm{imp}\frac{2\pi}{\hbar}\frac{e^4}{\epsilon_0^2}\frac{2-d_\mathrm{max}^2}{16\pi k^2} \label{eq:scat_coulomb}
~~~~\text{for} k_F \gg q_s,
\end{align}
and 
\begin{align}
    \frac{1}{\tau_{\bm{k}_F}} =
    n_\mathrm{imp}\frac{2\pi}{\hbar}\frac{e^4}{\epsilon_0^2}\frac{2-d_\mathrm{max}^2}{8\pi q_s^2}~~~~\text{for} k_F \ll q_s.  \label{eq:qs}
\end{align}
%
We consider the limit of $k_F \gg q_s$, where there is no screening effect. (The case of $k_F \ll q_s$ is in the main text.)
We calculate the response functions $L_{ij}$ with eq.~(\ref{eq:scat_coulomb}) changing the chemical potential $\mu$ and $d_\mathrm{max}$, as shown in Figure~\ref{Sfig1}. If we consider the ratio of $PF(d_\mathrm{max})$ and $PF(d_\mathrm{max}=0)$, one get
\begin{eqnarray}
PF(d_\mathrm{max})=\frac{2}{2-d_\mathrm{max}^2} PF(d_\mathrm{max}=0).
\end{eqnarray}
This is the same result with the case of $k_F\ll q_s$.

\begin{figure}[t]
\includegraphics[width=170mm]{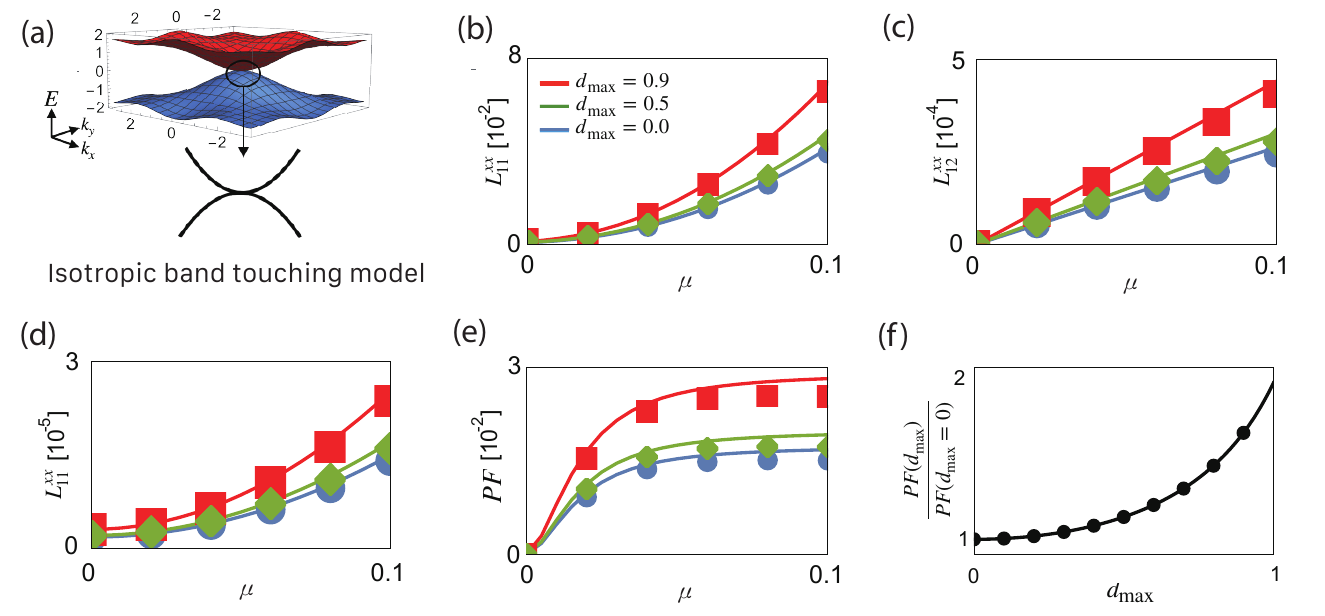} 
\caption{
(a) Band structure of eq.~(22) with $d_\mathrm{max}=0.9$ in the main text. (b-e) Chemical potential $\mu$ dependence of (b) $L_{11}$, (c) $L_{12}$, (d) $L_{22}$ and (e) $PF$ for $d_\mathrm{max}=0$(blue), 0.5(green) and 1(red). (f) represents the ratio between $PF(d_\mathrm{max})$ and $PF(0)$. The solid lines represent the results from the isotropic quadratic touching model in Eq.~(18) in the main text with $m_+^{-1}=-m_-^{-1}=1$
The discrete plots represent the results that are calculated from the lattice model Eq.~(22) in the main text. In this calculation, we set the parameters $T=0.01$, $k_B =\hbar=\gamma=u_0=e=\epsilon=1$, and consider the Coulomb-type impurity potential with $q_s \ll k_F$. \label{Sfig1}}
\end{figure}

\end{document}